\documentclass[twocolumn,showpacs,preprintnumbers,amsmath,amssymb]{revtex4}

\usepackage{graphicx}
\usepackage{bm}

\usepackage{amsthm}

\newcommand{\ep}{\epsilon}

\newcommand{\glog}{\Lambda}
\newcommand{\gexp}{{\cal E}}

\theoremstyle{definition}

\theoremstyle{remark}

\begin{document}

\title{What do generalized entropies look like? An axiomatic approach for complex, non-ergodic systems}

\author{Stefan Thurner$^{1,2}$ and Rudolf Hanel$^{1}$}

\affiliation{
$^1$ Section for Science of Complex Systems, Medical University of Vienna, Spitalgasse 23, A-1090, Austria\\
$^2$  Santa Fe Institute, 1399 Hyde Park Road, Santa Fe, NM 87501, USA
} 

\begin{abstract}
Shannon and Khinchin showed that assuming four information theoretic axioms the entropy must be of Boltzmann-Gibbs  type, 
$S=-\sum_i p_i \log p_i$. Here we note that in physical systems one of these axioms may be violated. For non-ergodic systems the so called
separation axiom  (Shannon-Khinchin axiom 4) will in general not be valid. We show that when this axiom is 
violated the entropy takes a more general form, $S_{c,d}\propto \sum_i ^W \Gamma(d+1, 1- c \log p_i)$, where $c$ and $d$ are scaling exponents
and $\Gamma(a,b)$ is the incomplete gamma function. 
The exponents $(c,d)$ define equivalence classes for  all interacting and non interacting systems
and  unambiguously characterize any statistical system in its thermodynamic limit. 
The proof is possible because of two newly discovered scaling laws which  
any entropic form has to fulfill, if the first three Shannon-Khinchin axioms hold. 
$(c,d)$ can be used to define equivalence classes of statistical systems. A series of known entropies  can be  classified in terms of these equivalence classes.
We show that the corresponding distribution functions are special forms of Lambert-${\cal W}$ exponentials containing  -- as  special cases  -- 
Boltzmann,  stretched exponential and Tsallis distributions (power-laws).  
In the derivation we assume trace form entropies, $S=\sum_i g(p_i)$, with $g$ some function, however more general entropic forms can 
be classified along the same scaling analysis. 
In this contribution we largely   follow the lines of thought presented in \cite{Hanel2011}. 
\end{abstract}


\maketitle
 
\section{Introduction}
Theorem number 2 in the seminal 1948 paper, {\em The Mathematical Theory of Communication} \cite{shannon},
by Claude Shannon, proves the existence of the one and only form of entropy, given that 
three fundamental requirements hold. A few years later A.I. Khinchin remarked  in his {\em Mathematical Foundations of Information Theory} \cite{kinchin_1}: 
``However, Shannon's treatment is not always sufficiently complete and mathematically correct so that, besides having to free the theory from practical  
details, in many instances I have amplified and changed both the statement of definitions and the statement of proofs of theorems.''
Khinchin adds a fourth axiom. The three fundamental requirements of Shannon, in the `amplified' version of Khinchin,  are known as the Shannon-Khinchin (SK) axioms. 
These axioms list the requirements needed for an entropy to be a reasonable measure of the `uncertainty'  about a finite probabilistic system. 
Khinchin further suggests to also use entropy as a measure of the information {\em gained} about a system when making an 'experiment', i.e. by observing a realization 
of the probabilistic system. 

Khinchin's first axiom states that for a system with $W$ potential outcomes (states) each of which is given by a probability $p_i\geq0$, with $\sum_{i=1}^W p_i=1$, 
the entropy $S(p_1, \cdots, p_W)$ as a measure of uncertainty about the system must take its maximum for the equi-distribution $p_i=1/W$, for all $i$.

Khinchin's second axiom (missing in \cite{shannon}) states that any  entropy should remain invariant under adding zero-probability states to the system, 
i.e. $S(p_1, \cdots, p_W)=S(p_1, \cdots, p_W,0)$. 

Khinchin's third axiom (separability axiom) finally makes a statement of the composition of two  finite probabilistic systems  $A$ and $B$. 
If the systems are independent of each other, entropy should be additive, meaning that the entropy of the combined system 
$A+B$ should be the sum of the individual systems, $S({A+B}) = S(A) + S(B)$. 
 If the two systems are dependent on each other, the entropy of the combined system,   
i.e. the information given by the realization of the two finite schemes $A$ and $B$, $S(A+B)$,  is equal to the 
information gained by a realization of system $A$,  $S(A)$, plus the mathematical expectation of information gained
 by a realization of system $B$, after the realization of system $A$, $S({A+B}) = S(A) + S|_A(B)$. 

Khinchin's fourth axiom is the requirement  that entropy is a continuous function of all its arguments $p_i$ and does not depend 
on anything else.

Given these axioms, the {\em Uniqueness theorem} \cite{kinchin_1} states that the one and only possible entropy is 
$S(p_1,\cdots , p_W) = -k \sum_{i=1}^{W}p_i\log p_i$, where $k$ is an arbitrary positive constant. The result is of course the same as Shannon's. 
We call the combination of 4 axioms  the Shannon-Khinchin (SK) axioms. 

From information theory now to physics, where systems may exist that violate the separability axiom. This might 
especially be the case for non-ergodic, complex systems exhibiting long-range  and strong interactions. 
Such complex systems may show extremely rich behavior in contrast to simple ones, such as gases.  
There exists some hope that it should be possible to understand such systems also on a thermodynamical basis, meaning 
that  a few measurable quantities would be sufficient to understand their macroscopic phenomena. 
If this  would be possible, through an equivalent to the  second law of thermodynamics, some appropriate entropy would enter  
as a fundamental concept relating the number of microstates in the system to  its macroscopic properties.  
Guided by this hope, a series of so called generalized entropies have been suggested over the past decades, 
see \cite{tsallis88, celia, kaniadakis,curado,expo_ent,ggent} and Table 1. 
These entropies have been designed for different purposes and have not been related to a fundamental origin. 
Here we ask how generalized entropies can look like if they fulfill some of  the Shannon-Khinchin axioms, 
but explicitly violate the separability axiom. We do this  axiomatically as first presented in \cite{Hanel2011}. 
By doing so we can relate a large class of generalized entropies to a single root. 

The reason why  this axiom is violated in some physical, biological or social systems is {\em broken ergodicity}, 
i.e. that not all regions in phase space are visited and many micro states are effectively `forbidden'. 
Entropy relates the number of micro states of a system to an {\em extensive} quantity, which plays the 
fundamental role in the systems thermodynamical description. Extensive means that if two initially isolated, i.e. sufficiently separated 
systems, $A$ and $B$, with $W_A$ and $W_B$ the respective numbers of states, are brought together, 
the entropy of the combined system 
$A+B$ is $S(W_{A+B}) = S(W_A) + S(W_B)$. 
$W_{A+B}$ is the number of states in the combined system $A+B$.
This is not to be confused with {\em additivity} which is the property that $S(W_A W_B) = S(W_A) + S(W_B)$. 
Both, extensivity and additivity coincide if  number of states in the combined system is $W_{A+B}=W_AW_B$.
Clearly,  for a non-interacting system Boltzmann-Gibbs-Shannon entropy, $S_{\rm BG}[p]= - \sum_i^W   p_i\ln p_i$,  is extensive {\em and} additive.
By 'non-interacting'  (short-range, ergodic, sufficiently mixing, Markovian, ...) systems we mean  $W_{A+B}=W_AW_B$. 
For interacting statistical systems the latter is in general not true; phase space is only partly visited and $W_{A+B} < W_AW_B$. In this case,   
an additive entropy  such as Boltzmann-Gibbs can no longer be  extensive and vice versa. 
To ensure extensivity of entropy, an entropic form should be found for the particular interacting statistical systems at hand. 
These entropic forms are called {\em generalized entropies}  and usually assume trace form \cite{tsallis88, celia, kaniadakis,curado,expo_ent,ggent}
\begin{equation}
 S_g[p]=\sum_{i=1}^W g(p_i) \quad ,
\label{S_g} 
\end{equation} 
$W$ being the number of states. 
Obviously not all generalized entropic forms are of this type. R\'enyi entropy e.g. is  
of the form $G(\sum_{i}^W g(p_i))$, with $G$ a monotonic function.
We use trace forms Eq. (\ref{S_g}) for simplicity. R\'enyi forms can be studied in exactly the same way 
as will be shown, however at more technical cost. 

Let us revisit the Shannon-Khinchin axioms in the light of generalized entropies of trace form   Eq. (\ref{S_g}). 
Specifically axioms SK1-SK3 (now re-ordered) have implications on  the functional form of $g$

\begin{itemize}
\item
SK1: The requirement that $S$ depends continuously on $p$  implies that $g$ is a continuous function.
\item
SK2: The requirement that the entropy is maximal for the equi-distribution $p_i=1/W$ 
(for all $i$)  implies that $g$ is a concave function.
\item 
SK3: The requirement that adding a zero-probability state to a system, $W+1$ with $p_{W+1}=0$,  
does not change the entropy, implies that $g(0)=0$. 
\item 
SK4 (separability axiom): The entropy of a system -- composed of sub-systems $A$ and $B$ -- equals the entropy of $A$ plus the expectation value of the entropy 
of $B$, conditional on $A$. Note that this also corresponds exactly to Markovian processes.
\end{itemize}
As mentioned, if SK1 to SK4 hold, the only possible entropy is the Boltzmann-Gibbs-Shannon entropy. 
We are now going to derive the extensive entropy when the separability axiom SK4 is violated. 
Obviously this entropy will be more general and should contain BG entropy as a special case. 

We now assume that axioms SK1, SK2, SK3  hold, i.e. we restrict ourselves to trace form entropies with $g$ continuous, concave 
and $g(0)=0$. These systems we call {\em admissible} systems.

This generalized entropy for  (large) admissible statistical systems (SK1-SK3 hold)   
is derived from two  hitherto unexplored fundamental scaling laws of extensive entropies. 
Both scaling laws are characterized by  exponents $c$ and $d$, respectively, which allow to uniquely define equivalence classes  of entropies,  
meaning that two entropies are equivalent in the thermodynamic limit 
if their exponents $(c,d)$ coincide. Each admissible system belongs to  one of these equivalence classes $(c,d)$, \cite{Hanel2011}. 

In terms of the exponents $(c,d)$ we show in the following that all generalized entropies  have the form 
$S_{c,d}\propto\sum_i ^W  \Gamma(d+1, 1- c \log p_i)$, 
with  $\Gamma(a,b)=\int_b^\infty dt\,t^{a-1}\exp(-t)$ the incomplete Gamma-function. 

Admissible systems when combined  with a maximum entropy principle show remarkably simple mathematical 
properties as recently demonstrated in \cite{Hanel2011_b}.

\begin{figure}[t]
 \begin{center}
 	\includegraphics[width=\columnwidth] {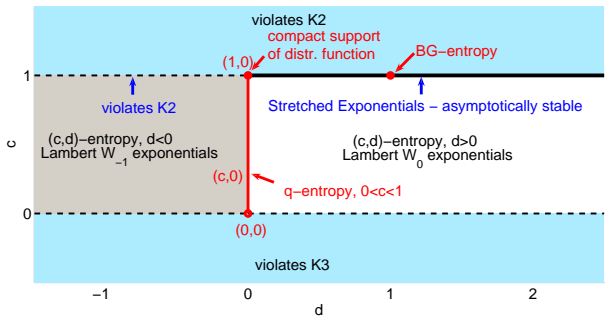}
 \end{center}
\caption{
Equivalence classes of entropies parametrized in the $(c,d)$-plane, with their associated distribution functions. 
BG entropy corresponds to $(1,1)$, Tsallis entropy to $(c,0)$, and entropies for stretched exponentials to $(1,d>0)$.
All entropies leading to distribution functions with compact support, 
belong to equivalence class $(1,0)$. An example are $S_q$ entropies with $q>1$ (using the maximum entropy principle with usual expectation values in the constraints \cite{ggent,HT_hilhorst}).
\label{classfig}
}
\end{figure}

\section{The scaling laws of  entropies}

We discuss two  -- a primary and a secondary -- scaling properties of generalized entropies of trace form and assume the validity of  the first 3 KS axioms. 
For equi-distribution $p_i=\frac 1W$ (for all $i$) obviously, $\sum_{i=1}^W g(p_i)= W g(\frac 1W)$. 

The {\em primary scaling law} is found from the relation 
\begin{equation}
 \frac{S_g(\lambda W)}{S_g(W)}=\lambda \frac{g( \frac{1}{\lambda W} )}{g(\frac 1W)}\quad . 
\label{Stog}
\end{equation}
We define a scaling function 
\begin{equation}
   f(z)\equiv\lim_{x\to 0}\frac{g(z x)}{g(x)}    \quad \quad  (0<z<1) \quad. 
\label{f_funct}
\end{equation}
This function $f$ for systems satisfying SK1-SK3, but violating SK4, 
can only be a power $f(z)=z^c$, with $0<c\leq 1$,  given $f$ being continuous. 
This mathematical fact is shown as a theorem in the appendix (Theorem 1). 
Inserting Eq. (\ref{f_funct}) in Eq. (\ref{Stog}) gives the primary asymptotic scaling law 
for entropies
\begin{equation}
 \lim_{W \to \infty }\frac{S_g(\lambda W)}{S_g(W)}=  \lambda ^{1-c} \quad . 
\label{Stog2}
\end{equation}

To identify the {\em secondary scaling law} start from the observation  that 
\begin{equation}
    \lim_{W\to \infty}   \frac{S(\lambda W)}{S(W)}  \lambda^{c-1} =1 \quad.  
\label{f_funct2}
\end{equation}
and substitute  $\lambda$ in Eq. (\ref{f_funct2}) by $\lambda \to W^a$.  We define $h_c(a)$
\begin{equation}
  h_c(a)\equiv    \lim_{W\to \infty}   \frac{S(W^{1+a})}{S(W)}  W^{a(c-1)}  = \lim_{x\to 0}  \frac{g(x^{1+a})}{ x^{ac}g(x)} \quad, 
\label{fr_funct}
\end{equation}
with $x\equiv1/W$. 
It can be  proved as a mathematical fact (Theorem 2 in the appendix) that 
\begin{equation}
  h_c(a)=(1+a)^d \qquad    (d \,\,\, {\rm constant}) \quad. 
\end{equation}
Remarkably, $h_c$ does not explicitly depend on $c$. $h_c(a)$ is an asymptotic property which is 
{\em independent}  of property Eq. (\ref{Stog2}). 
Note that if $c=1$, concavity of $g$ implies $d\geq 0$.  
In principle this scheme can be iterated to higher orders,
see appendix. 

\section{Derivation of entropy}

We now ask which families of entropies, i.e. functions $g_{c,d}$
fulfill the primary and the secondary scaling law. 
A particularly simple choice is 
\begin{equation}
g_{c,d,r}(x) = r A^{-d}e^{A} \, \Gamma \left(1+d\,,\,A-c\ln x \right)-rcx  \quad, 
\label{gent}
\end{equation}
where $A=\frac{cdr}{1-(1-c)r}$
and $r$ is an arbitrary constant $r>0$ (see below). 
For all choices of $r$ the function 
$g_{c,d,r}$ is a representative of the class $(c,d)$.  
This allows to choose $r$ as a suitable function of $c$ and $d$. 
For example choose $r=(1-c +cd)^{-1}$, so that $A=1$, and 
\begin{equation}
S_{c,d}[p] = \frac{e \sum_i^W \Gamma \left(1+d\,,\,1-c\ln p_i \right)}{1-c+cd} - \frac{c}{1-c+cd}  \quad. 
\label{gent2}
\end{equation}
It can be easily verified that $S_{c,d}$ has the correct asymptotic properties, 
see Theorem 3  in the appendix.

\subsection{Special cases of entropic equivalence classes}
Let us look at some specific equivalence classes $(c,d)$
\begin{itemize}
\item Boltzmann-Gibbs entropy belongs to the $(c,d)=(1,1)$ class. One gets from Eq. (\ref{gent})  
 	\begin{equation}
  	S_{1,1}[p]= \sum_i g_{1,1}(p_i)=   -\sum_i p_i\ln p_i +1  
  	\quad .
 	\end{equation}

\item Tsallis entropy belongs to the  $(c,d)=(c,0)$ class. From Eq. (\ref{gent}) and the choice $r=1/(1-c)$ (see below)
	we get
 	\begin{equation}
		\begin{array}{lcl}
	 		S_{c,0}[p] = \sum_i g_{c,0}(p_i)=  \frac{1-\sum_i p_i^c}{c-1} +1 \, . 
		\end{array}
 	\end{equation}
	Note, that although the {\em pointwise} limit $c\to 1$ of Tsallis entropy yields  BG entropy, the asymptotic properties $(c,0)$ do {\em not} 
	change continuously to $(1,1)$ in this limit! In other words the thermodynamic limit and the limit $c\to 1$ do not commute.

\item The entropy related to  stretched exponentials \cite{celia} belongs  to the $(c,d)=(1,d)$ classes, see Table 1.
	As a specific example  we compute the $(c,d)=(1,2)$ case,  
	\begin{equation}
  		S_{1,2}[p]= 2  \left(1-\sum_i p_i \ln p_i  \right) + \frac{1}{2}\sum_i p_i\left(\ln p_i  \right)^2, 
	\label{c1d2}
 	\end{equation}
 	leading to a superposition of two entropy terms, the asymptotic behavior being dominated by the second.  	
	
\item All entropies associated to distributions with  compact support belong to $(c,d)=(1,0)$. 	
Clearly, all of these have the same (trivial) asymptotic behavior. 
\end{itemize}
Other entropies which are special cases of our scheme are found in Table 1. 

Inversely, for any given entropy we are now in the remarkable position to characterize {\em all} 
large SK1-SK3 systems by a pair of two exponents $(c,d)$, i.e. 
their scaling functions $f$ and $h_c$. See Fig. \ref{classfig}.
For example, for $g_{\rm BG}(x)=-x\ln(x)$ we have $f(z)=z$, i.e.  $c=1$, 
and $h_{c}(a)=1+a$, i.e.  $d=1$.
$S_{\rm BG}$ therefore belongs to the universality class $(c,d)=(1,1)$.  
For $g_{q}(x)= (x-x^q)/(1-q)$ (Tsallis entropy) and $0<q<1$ one finds 
$f(z)=z^q$, i.e.  $c=q$ and $h_{c}(a)=1$, i.e.  $d=0$, 
and Tsallis entropy, $S_{q}$, belongs to the universality class $(c,d)=(q,0)$. 
Other examples are listed in Table 1. 

The universality classes $(c,d)$ are equivalence classes with the 
equivalence relation given by:  
$g_{\alpha} \equiv g_{\beta} \Leftrightarrow c_\alpha=c_\beta$ and $d_\alpha=d_\beta$.
This relation partitions the space of all admissible $g$ into equivalence classes completely specified by the pair $(c,d)$.  

\begin{figure}[t]
 \begin{center}
	\includegraphics[width=0.7\columnwidth] {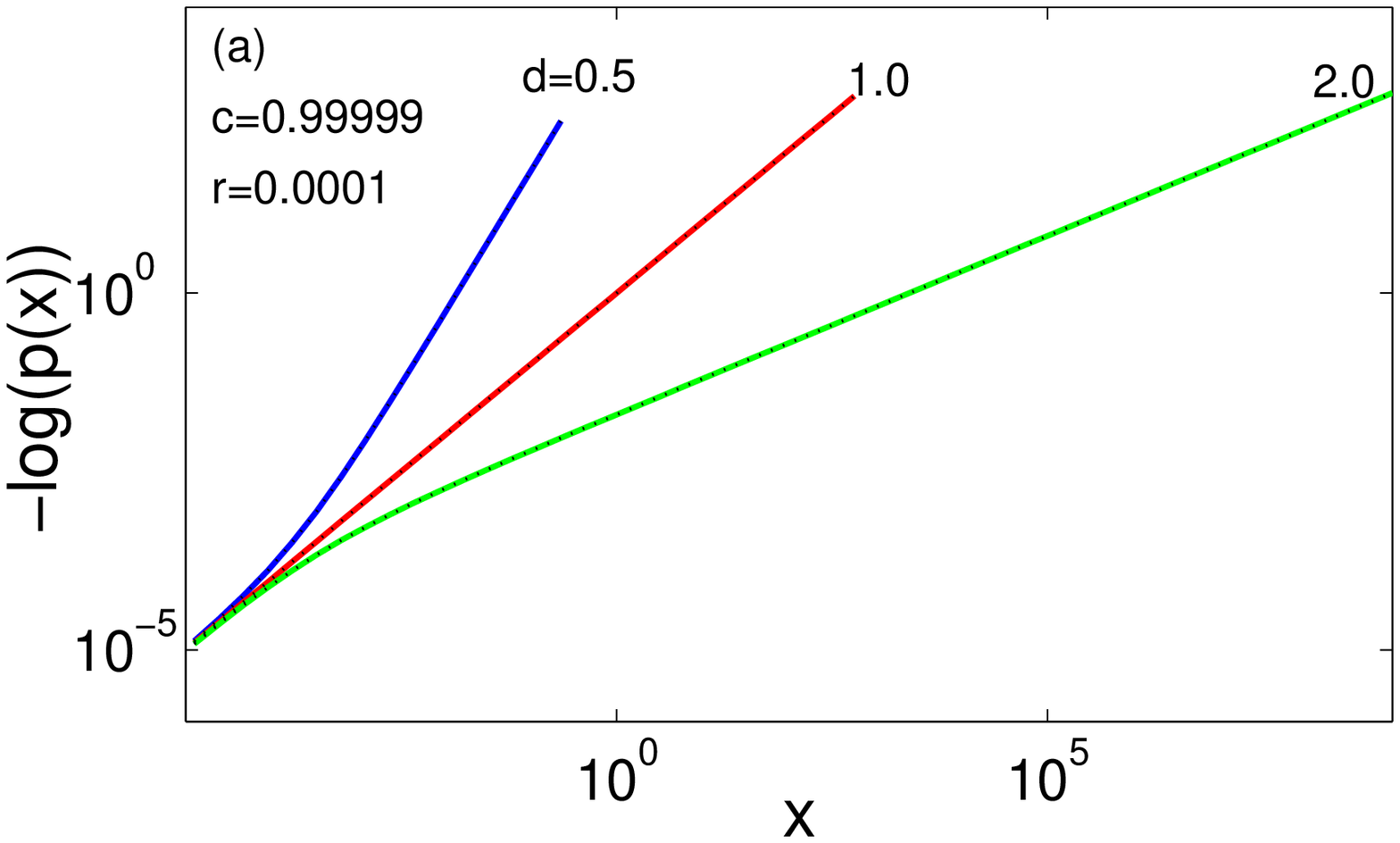}
	\includegraphics[width=0.7\columnwidth] {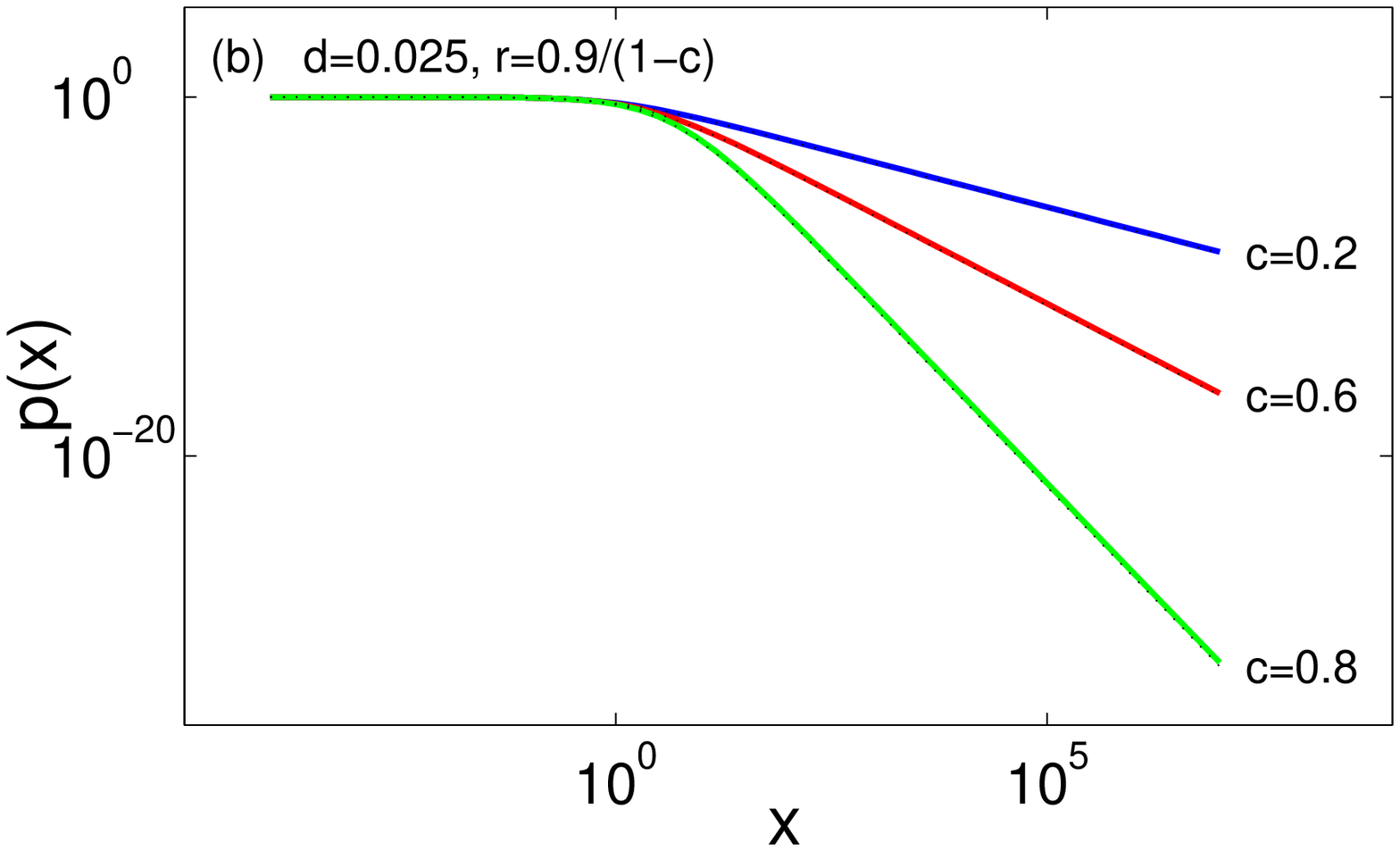}
	\includegraphics[width=0.7\columnwidth] {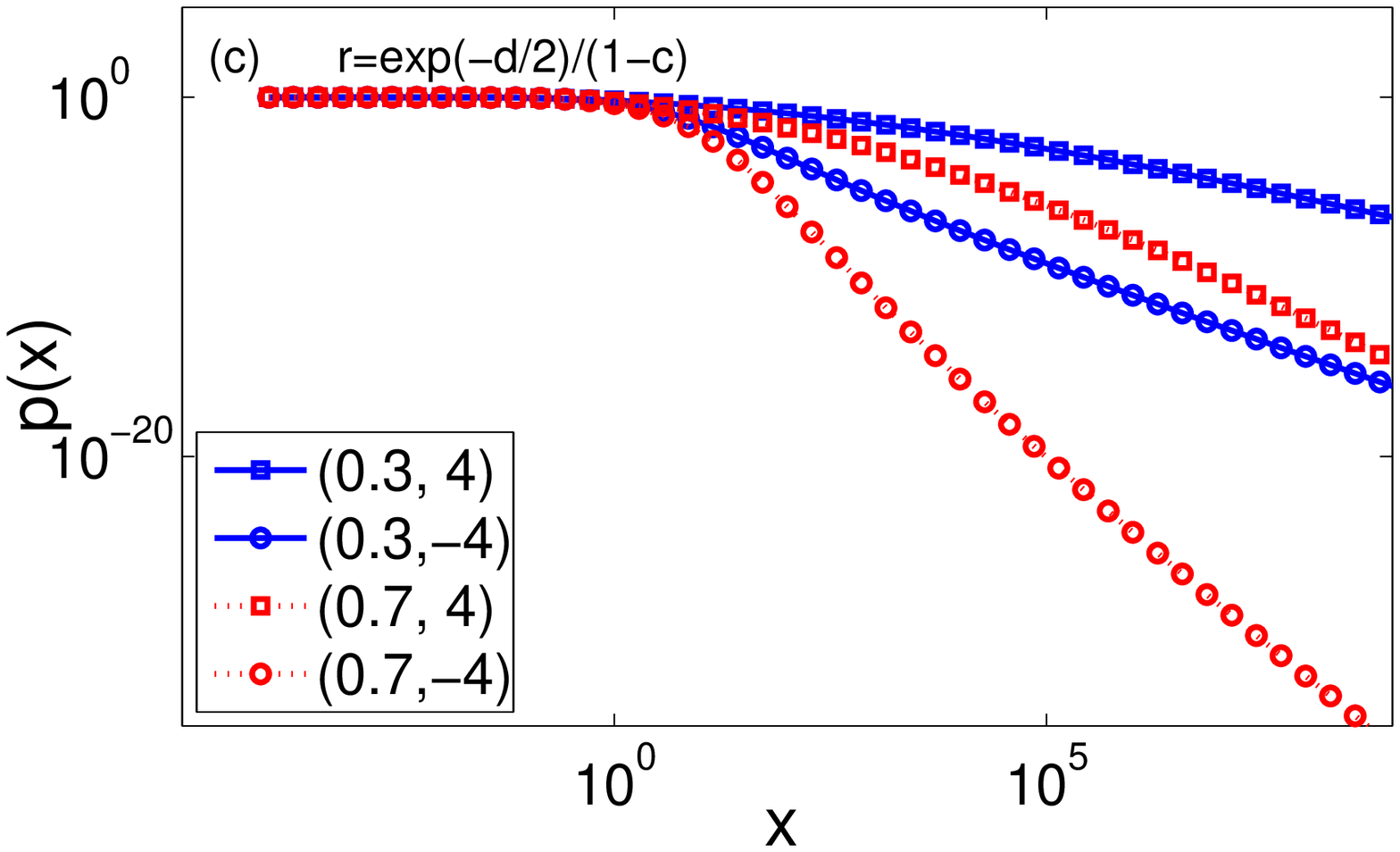}
 \end{center}
\caption{
Distribution functions based on the  Lambert exponential, $p(x)=\gexp_{c,d,r}(-x)$, for various $(c,d)$ values.
(a) Stretched exponential limit $c\to 1$.  It includes the Boltzmann distribution for $d=1$.
(b) $d\to 0$ limit -- i.e. the $q$-exponential limit. In (a) and (b) the black dashed lines represent the stretched exponential ($c=1$) or $q$-exponential ($d=0$)
limit functions.
(c) General case for distribution functions for several values of $(c,d)$ different from the limits $c\sim 1$ or $d\sim 0$. They must not 
be confused with power-laws. 
\label{MECfig}
}
\end{figure}

\begin{table*}[t]
\caption{
Several trace form entropies  $S=\sum_i^W g(p_i)$, for which  SK1-SK3 hold. They are  special cases of the entropy given in Eq. (\ref{gent}) and their  asymptotic behavior 
is uniquely determined by $c$ and $d$. 
It can be seen immediately that $S_{q>1}$, $S_{b}$ and $S_{E}$ are asymptotically identical; so are $S_{q<1}$ and $S_{\kappa}$, as well as $S_{\eta}$ and $S_{\gamma}$.  
}
\centering
\begin{tabular}{ll|c|c|l}
entropy &  & $c$ & $d$ & reference \\ \hline
$S_{c,d} = er \sum_i \Gamma(d+1, 1-c\ln p_i) - cr $ 	& $(r=(1-c+cd)^{-1})$								& $c$ 		& $d$  			&  \\  \hline
$S_{BG} =  \sum_i p_i  \ln  (1/p_i)    $ 			&												& $1$ 		& $1$  			&   \cite{kinchin_1} \\  \hline
$S_{q<1}(p) = \frac{1-\sum{p_i^q}}{q-1}$ 			& $(q<1)$ 										& $c=q<1$ 	& $0$  			&  \cite{tsallis88} \\  \hline
$S_{\kappa}(p) = - \sum_i p_i \frac{p_i^{\kappa}  - p_i^{-\kappa}  }{2\kappa}  $ 	 &	($0<\kappa \leq1$)						& $c=1-\kappa$ & $0$  			&   \cite{kaniadakis} \\  \hline
$S_{q>1}(p) = \frac{1-\sum{p_i^q}}{q-1}$ 			& $(q>1)$ 										& $1$ 		& $0$  			&   \cite{tsallis88} \\  \hline
$S_{b}(p) = \sum_i (1-e^{-bp_i})   + e^-b -1$ 		& $(b>0)$ 										& $1$ 		& $0$  			&   \cite{curado} \\  \hline
$S_{E}(p) =  \sum_i p_i (1-e^{\frac{p_i -1}{p_i}})    $  &												& $1$ 		& $0 $  			&   \cite{expo_ent} \\  \hline
$S_{\eta}(p) = \sum_i  \Gamma(\frac{\eta+1}{\eta},-\ln p_i) - p_i \Gamma (\frac{\eta+1}{\eta})$  &$ (\eta>0)$ 		& $1$ 		& $d=\frac{1}{\eta}$  	&   \cite{celia} \\  \hline
$S_{\gamma}(p) =  \sum_i p_i \ln ^{1  / \gamma} (1/p_i)    $ 	&										& $1$ 		& $d =1/\gamma $  	&   \cite{TsallisBook_2009}, footnote 11, page 60 \\  \hline
$S_{\beta}(p) =  \sum_i p_i^{\beta} \ln  (1/p_i)    $ 	&												& $c=\beta$  	& $1$  	&    \cite{shafee} \\  \hline

\end{tabular}
\end{table*}

\section{Distribution functions}

Distribution functions associated with  our  $\Gamma$-entropy, Eq. (\ref{gent2}), 
can be derived from so-called generalized logarithms of the entropy. Under the 
maximum entropy principle (given ordinary constraints) the inverse functions of these logarithms, $\gexp=\glog^{-1}$, 
are the distribution functions, 
$p(\epsilon) = \gexp_{c,d,r}(-\epsilon)$. 
Following \cite{ggent,HT_hilhorst}  the generalized logarithm $\glog$  is found in closed form
\begin{equation}
\glog_{c,d,r}(x) =   r-   r \, x^{c-1} \,  \left[ 1- \frac{1-(1-c)r}{rd} \ln x  \right]^{d}  \,,
\label{glog}
\end{equation}
and its inverse function is  
\begin{equation}
\gexp_{c,d,r}(x)=  e^{    - \frac{d}{1-c}  \left[ {\cal W}_k \left( B (1-x/r )^{ \frac{1}{d} } \right)  -  {\cal W}_k(B)  \right]  }\, ,
\label{gexp}
\end{equation}
with the constant $B\equiv \frac{(1-c)r}{1-(1-c)r} \exp \left(  \frac{(1-c)r}{1-(1-c)r} \right) $. 
The function ${\cal W}_k$ is the $k$'th branch of the Lambert-${\cal W}$ function which -- as a solution to the equation
$x={\cal W}(x)\exp({\cal W}(x))$ -- has only two real solutions $W_k$, the branch $k=0$ and branch $k=-1$. 
Branch $k=0$ covers  the classes for $d\geq 0$, branch $k=-1$ those for $d<0$.

\subsection{Special cases of distribution functions}

It is easy to verify that the class $(c,d)=(1,1)$ leads to Boltzmann distributions, and the class $(c,d)=(c,0)$ yields 
power-laws, or more precisely, Tsallis distributions i.e.  $q$-exponentials. 
All classes associated with  $(c,d)=(1,d)$, for $d>0$ are associated with stretched exponential distributions.   
Expanding the $k=0$ branch of the Lambert-${\cal W}$ function  $W_0(x)\sim x-x^2+\dots$ for $1\gg|x|$, 
the limit $c\to 1$ is seen to be a stretched exponential 
\begin{equation}
	\lim_{c\to 1}\gexp_{c,d,r}(x)=e^{-dr\left [\left(1-\frac{x}{r}\right)^{\frac{1}{d}}-1\right]}\quad. 
\label{strexp}
\end{equation}
Clearly,  $r$ does not effect its asymptotic properties (tail of the distribution), but can be used to incorporate  
finite size properties of the distribution function for small $x$.  
Examples of all of the above distribution functions  are shown in Fig. \ref{MECfig}. 

\subsection{Finite size effects and the parameter $r$}

In Eq. (\ref{gent2}) we chose $r=(1-c+cd)^{-1}$. This is not the most general case 
and the only  limitations on $r$ -- if one requires the generalized logarithms 
to have the usual properties $\glog(1)=0$ and $\glog'(1)=1$ to hold  -- are 
\begin{equation}
\begin{array}{ll}
d>0:& r<\frac{1}{1-c}\quad,\\
d=0:& r=\frac{1}{1-c}\quad,\\
d<0:& r>\frac{1}{1-c}\quad.
\end{array}
\label{rrange}
\end{equation}
Every choice of $r$ gives a representative of the equivalence class $(c,d)$,
i.e. $r$ has no effect on the thermodynamic limit and therefore can be used to encode potential  finite-size characteristics 
of a system at hand. In case of no such effects, practical choices are 
$r=(1-c+cd)^{-1}$ for $d>0$, and $r=\exp(-d)/(1-c)$ for $d<0$.

\section{A physical system as an example}

From axiomatic derivations  no physical relevance can be inferred.  To demonstrate the 
applicability of the proposed classification scheme to  concrete physical systems, 
consider over-damped interacting particles moving in a narrow channel, 
\begin{equation} 
\mu \vec v_i = \sum_{j\neq i} \vec J (\vec r_i-r_j) + \vec F(\vec r_i) + \eta(\vec r_i,t) \quad,
 \end{equation}
where $v_i$ is  velocity of $i$th particle, $\mu$  viscosity, $F$ an external force,  
$\vec J (\vec r)$
a linear repulsive particle-particle interaction, 
$\eta$ uncorrelated thermal noise with $\langle \eta \rangle=0$ and $\langle \eta^2 \rangle=\frac{kT}{\mu}$, with 
$\lambda$ a characteristic length of the pair interaction.  For details see \cite{AndradeNobreCurado},  
where the non-linear Fokker-Planck equation for the spatial distribution $\rho(x)$ of the particles was solved. 
It was shown that  the corresponding entropy is a superposition of BG and  Tsallis entropy with $q=2$. 
Note the similarity to our example in Eq. (\ref{c1d2}) where a similar superposition emerges naturally. 
At high temperatures the BG contribution dominates i.e. particles diffuse normally, while at zero temperature the 
system is governed by Tsallis entropy with a non-Gaussian parabolic diffusion profile of compact support. 
The concrete stationary distribution functions given in \cite{AndradeNobreCurado} can be used to demonstrate that the system is
found either in the asymptotic equivalence class $(c,d)=(1,1)$ (BG entropy) or in $(c,d)=(1,0)$ (compact support entropies) 
depending on the temperature of the heat bath.
Expressing the integral by a discrete sum  the entropy given in \cite{AndradeNobreCurado} 
can be used to demonstrate the existence of the two sets of classes. 
This shows that the classification is applicable for concrete physical systems and suggests further 
that the equivalence class $(c,d)$ may even depend on macro variables of the system such as the temperature
of the heat bath.

\section{R\'enyi-type entropies}

R\'enyi entropy is obtained by relaxing SK4 to the pure (unconditional) additivity condition. 
Following the same scaling idea for R\'enyi-type entropies,  $S=G(\sum_{i=1}^W g(p_i))$, with $G$ and $g$ some functions, one gets 
\begin{equation}
\lim_{W\to\infty} \frac{S(\lambda W)}{S(W)}= \lim_{s\to\infty} \frac{ G \left( \lambda f_g(\lambda^{-1})s \right)  }{G(s)} \quad ,
\end{equation} 
where $f_g(z)=\lim_{x\to 0} g(zx)/g(x)$. 
The expression $f_G(s)\equiv \lim_{s} G(s y)/G(s)$, 
provides the starting point for deeper analysis which now gets more involved. 
In particular, for R\'enyi entropy with $G(x) \equiv \ln(x)/(1-\alpha)$ and $g(x)\equiv x^{\alpha}$, 
the asymptotic properties yield the class $(c,d)=(1,1)$, (BG entropy) meaning that R\'enyi entropy is additive. 
However, in contrast to the trace form entropies used above, 
R\'enyi entropy can be shown to be {\em not} Lesche stable, as was observed before \cite{lesche,Abe_2002, Arimitsu,Kaniadakis_2004,HT_robustness_1}. 
All of the $S=\sum_i^W g(p_i)$ entropies can be shown to be Lesche stable, see Theorem 4 in the appendix. 

\section{Discussion}

We studied the scaling laws of trace form entropies which are constrained by the first three Shannon-Khinchin axioms in the thermodynamic limit. 
In analogy to critical exponents these laws are characterized by two scaling  exponents $(c,d)$, 
which  define equivalence relations on  entropic forms. 
We showed that a single entropic form -- parametrized by the two exponents --  
covers all {\em admissible} systems (Shannon-Khinchin axioms 1-3 hold, 4 is violated). 
In other words every statistical system has its pair of unique exponents in the large size limit, its entropy is then given by 
$S_{c,d} \sim \sum_i^W  \Gamma \left(1+d\,,\,1-c\ln p_i \right)$. 

The exponents for BG systems are $(c,d)=(1,1)$, systems characterized by stretched exponentials  
belong to the class $(c,d)=(1,d)$, and Tsallis systems have $(c,d)=(q,0)$. 
In the context of a maximum entropy principle, the associated distribution functions of {\em all} systems $(c,d)$ are  shown 
to belong to a class of  exponentials  involving Lambert-${\cal W}$ functions, given in Eq. (\ref{gexp}). 
There are no other options for tails in distribution functions other than these.  

The equivalence classes characterized by the exponents $(c,d)$ form {\em basins of asymptotic equivalence}.
In general these basins  characterize interacting statistical (non-additive) systems.
There exists an analogy between these basins of asymptotic {\em equivalence} and
the {\em basin of attraction} of weakly interacting, uncorrelated systems subject to the law of large numbers, i.e. the central limit theorem. 
Any system within a given equivalence class may show 
individual characteristics as long as it  is small. Systems belonging to the same class will start behaving
similarly as they become larger and in the thermodynamic limit they become identical.  
Distribution functions converge to those uniquely determined by $(c,d)$.

Our framework shows that for non-interacting systems $c=1$.
Setting $\lambda=W_B$ in Eq. (\ref{Stog}) and Eq. (\ref{f_funct}), immediately implies $S(W_AW_B)/S(W_A)\sim  W_B^{1-c}$.
This means that if  for such a system it would be true that $c\neq1$,  
then adding only a few independent states to a system would explosively change its entropy and extensivity would be strongly violated.
A further interesting feature of admissible systems is that they all are  what has been called {\em Lesche stable}. 
As a practical note Lesche stability corresponds one-to-one to the continuity of the scaling function $f$ 
and can now be checked by a trivial verification of this property (Eq. (\ref{f_funct})). The proof is given in Theorem 4 in the appendix.

Finally we remark that the classification scheme for generalized entropic 
forms of type $S=\sum_i g(p_i)$ can be  extended to entropies of e.g. R\'enyi type, i.e. $S=G(\sum_{i} g(p_i))$. 
We demonstrated that generalized entropies can be applied to actual physical systems. We hypothesize that many complex statistical 
systems are indeed admissible systems  of equivalence classes $(c,d)$, with $0<c<1$.


\section{Appendix}

{\bf Theorem 1:} 
Let $g$ be a continuous, concave function on $[0,1]$ with
$g(0)=0$ and 
let $f(z)=\lim_{x\to 0^+}g(zx)/g(x)$ be continuous, then
$f$ is of the form $f(z)=z^c$ with $c\in(0,1]$. 

\begin{proof}
Note that $f(ab)=\lim_{x\to 0}g(ab x)/g(x)=\\ \lim_{x\to 0}(g(ab x)/g(b x))(g(b x)/g(x))=f(a)f(b)$. 
All pathological solutions are excluded by the requirement that $f$ is continuous.
So $f(ab)=f(a)f(b)$ implies that $f(z)=z^c$ is the only possible solution of this equation. 
Further, since $g(0)=0$, also $\lim_{x\to 0}g(0 x)/g(x)=0$, and it follows that $f(0)=0$.
This necessarily implies that $c>0$. $f(z)=z^c$ also has to be concave since $g(z x)/g(x)$ is
concave in $z$ for arbitrarily small, fixed $x>0$. Therefore $c\leq 1$. 
\end{proof}

{\bf Theorem 2:}
Let $g$ be like in Theorem 1 and let $f(z)=z^c$ then
$h_c$ in Eq. (\ref{fr_funct}) is a constant of the form 
$h_c(a)=(1+a)^d$ for some constant $d$.

\begin{proof}
We can determine $h_c(a)$ again by a similar trick as we have used for $f$.
\begin{equation} 
\begin{array}{ll} 
h_c(a)&= \lim_{x\to 0} \frac{g(x^{a+1})}{x^{ac}g(x)}\\
&=\frac{g\left((x^b)^{\left(\frac{a+1}{b}-1\right)+1}\right)}{(x^b)^{\left(\frac{a+1}{b}-1\right)c}g(x^b)}\frac{g(x^{b})}{x^{(b-1)c}g(x)}\\ \nonumber
&=h_c\left(\frac{a+1}{b}-1\right)h_c\left(b-1\right)\quad,
\end{array}
\end{equation}
for some constant $b$.
By a simple transformation of variables, $a=bb'-1$, one gets 
$h_c(bb'-1)=h_c(b-1)h_c(b'-1)$. 
Setting $H(x)=h_c(x-1)$ one again gets $H(bb')=H(b)H(b')$.
So $H(x)=x^d$ for some constant $d$ and consequently 
$h_c(a)$ is of the form $(1+a)^d$.
\end{proof}

{\bf Theorem 3:} 
The entropy based on $g_{c,d,r}$, Eq. (\ref{gent}),  has the desired
asymptotic properties. 

\begin{proof}
Let $g$ be like in Theorem 1,  i.e. let $f(z)=z^c$ with $0<c\leq 1$, then
\begin{equation}
\lim_{x\to 0^+} \frac{g'(x)}{\frac{1}{x}g(x)}=c\,.
\label{prehosp}
\end{equation}
Consider 
\begin{equation}
\begin{array}{lcl}
\lim_{x\to 0^+} \frac{\frac{g(x)-g(zx)}{(1-z)x}}{\frac{1}{x}g(x)}&=&\frac{1}{1-z}\left(\frac{g(x)-g(zx)}{g(x)}\right)
=\frac{z^c-1}{z-1}\,.
\nonumber
\end{array}
\end{equation}
Taking the limit $z\to 1$ on both sides completes the first part of the proof.
Further, two functions $g_{A}$ and $g_{B}$ 
generate equivalent entropic forms
if $\lim_{x\to 0^+}g_A(x)/g_B(x)=\phi$ and $0<\phi<\infty$. This clearly is true since
\begin{equation}
\begin{array}{lcl}
\lim_{x\to 0^+}\frac{g_A(zx)}{g_A(x)}&=&\frac{g_A(zx)}{g_B(zx)}\frac{g_B(x)}{g_A(x)}\frac{g_B(zx)}{g_B(x)}\\
&=&\phi\phi^{-1}\frac{g_B(zx)}{g_B(x)}\\
&=&\lim_{x\to 0^+}\frac{g_B(zx)}{g_B(x)}\,.
\nonumber
\end{array}
\end{equation}
By an analogous argument the same result can be obtained for the second asymptotic property, Eq. (\ref{fr_funct}). 
A simple lemma is that given $g_B(x)=ag_A(bx)$, for some suitable constants $a$ and $b$,
then $g_B$ and $g_A$ are equivalent.

A lemma, following from Eq. (\ref{prehosp}) is that 
\begin{equation}
\lim_{x\to 0^+} \frac{g_A(x)}{g_B(x)}=\lim_{x\to 0^+}\frac{g'_A(x)}{g'_B(x)}\,,
\nonumber
\end{equation}
which is just the rule of L'Hospital shown to hold for the considered families of functions $g$.
This is true since, either $\lim_{x\to 0^+} g_A(x)/g_B(x)=\phi$ with $0<\phi<\infty$ and $c_A=c_B$,
i.e. $g_A$ and $g_B$ are equivalent, or $g_A$ and $g_B$ are inequivalent, i.e.
$c_A\neq c_B$ but $\phi=0$ or $\phi \to \infty$.  

So if one can find a function $g_{\rm test}$, having the desired asymptotic exponents $c$ and $d$, it suffices to show that
$0<-\lim_{x\to0^+}\Lambda_{c,d,r}(x)/g'_{\rm test}(x)<\infty$, where $\Lambda_{c,d,r}$ is the generalized logarithm
Eq. (\ref{glog}) associated with the generalized entropy Eq. (\ref{gent2}).
The test function $g_{\rm test}(x)=x^c\log(1/x)^d$ is of class $(c,d)$, as can be verified easily.
Unfortunately $g_{\rm test}$ can not be used to define the generalized 
entropy due to several technicalities. In particular $g_{\rm test}$ lacks concavity around $x\sim 1$ 
for a considerable range of $(c,d)$ values, which then makes it impossible to define
proper generalized logarithms and generalized exponential functions on the entire interval $x\in[0,1]$. 
However, we only need the asymptotic properties of $g_{\rm test}$ and for 
$x\sim 0$ the function $g_{\rm test}$ does not violate concavity or any other required condition.
The first derivative is $g'_{\rm test}(x)=x^{c-1}\log(1/x)^{d-1}(c\log(1/x)-d)$. 
With this we finally get 
\begin{equation}
\lim_{x\to 0^+}\frac{\Lambda_{c,d,r}(x)}{g'_{\rm test}(x)}=\frac{r-D^{-d}\left(\frac{x}{z}\right)^{c-1}\left[\log\left(\frac{z}{x}\right)\right]^{d}}  {x^{c-1}\log(\frac 1x)^{d-1}(c\log \frac 1x -d)}
=-\frac{z^{1-c}}{cD^d}\,.
\end{equation}
Since $0<\frac{z^{1-c}}{cD^d}<\infty$ this 
proves that the Gamma-entropy $g_{c,d,r}$, Eq. (10), represents the equivalence classes $(c,d)$. 
\end{proof}

{\bf  Lesche stability of trace form entropies} 

The Lesche stability criterion is a uniform-equi-continuity property of functionals $S[p]$ on
families of probability functions $\{p^{(W)}\}_{W=1}^{\infty}$ where
$p^{(W)}=\{p_i^{W}\}_{i=1}^W$. The criterion is phrased as follows:\\

Let $p^{(W)}$ and $q^{(W)}$ be probabilities on $W$ states. An entropic form $S$ is Lesche stable 
if for all $\ep>0$ and all $W$ there is a $\delta>0$ such that
\begin{equation}
||p^{(W)}-q^{(W)}||_1<\delta \Rightarrow |S[p^{(W)}]-S[q^{(W)}]|<\ep \hat S(W)\quad,
\end{equation}
where $\hat S(W)$ is again the maximal possible entropy for $W$ states.\\

We characterize Lesche stability on the class of our generalized entropic forms in terms of continuity of $f$ in \\

{\bf Theorem 4:} 
Let $p_i\geq0$ be a probability and $W$ the number of states $i$. 
Let $g$ be  a concave, continuous function on $[0,1]$, 
continuously differentiable on the semi-open interval $(0,1]$ and $g(0)=0$.
The entropic form $S_g[p]=\sum_{i=1}^{W} g(p_i)$ is 
Lesche stable iff the function $f(z)=\lim_{x\to 0}g(z x)/g(x)$ is continuous on $z\in[0,1]$. 

\begin{proof} 
SK2  states that maximal entropy is given by  $\hat S_g(W)=Wg(1/W)$.
We identify the worst case scenario for $|S_g[p]-S_g[q]|$, where $p$ and $q$ are probabilities on the 
$W$ states. For this maximize $G[p,q]=|S_g[p]-S_g[q]|-\alpha(\sum_i p_i-1)-\beta(\sum_i q_i-1)-\gamma(\sum_i|p_i-q_i|-\delta)$, 
where $\alpha$, $\beta$ and $\gamma$ are Lagrange multipliers.
Without loss of generality assume that $S_g[p]>S_g[q]$. Thus  condition $\partial G/\partial p_i=0$ gives
$g'(p_i)+\gamma \, {\rm sign}(p_i-q_i)-\alpha=0$, where $g'$ is the derivative of $g$ and  ${\rm sign}$ is the sign function.
Similarly, $\partial G/\partial q_i=0$ leads to $g'(q_i)+\gamma \, {\rm sign}(p_i-q_i)+\beta=0$. From this we see that both 
$p$ and $q$ can only possess two values $p_+$, $p_-$ and $q_+$ and $q_-$, where one can assume (without loss of generality)
that $p_+>q_-$ and $q_+>p_-$. We can now assume that for $w$ indices $i$ $p_+=p_i>q_i=q_-$ and for $W-w$ indices $j$
$p_-=p_j<q_j=q_+$ where $w$ may range from $1$ to $W-1$. This leads to seven equations 
\begin{equation}
\begin{array}{lll}
wp_++(W-w)p_-=1 &,& g'(p_+)+\gamma-\alpha=0 \\ 
wq_-+(W-w)q_+=1 &,& g'(p_-)-\gamma-\alpha=0 \\ 
w(p_+-q_-) -    &,& g'(p_+)+\gamma+\beta=0 \\  
-(W-w)(p_--q_+)=\delta &,& g'(p_+)-\gamma+\beta=0  
\end{array}
\end{equation}
which allow to express $p_-$, $q_-$, and $q_+$ in terms of $p_+$ 
\begin{eqnarray}
p_-&=& \frac{1-w p_+}{W-w} \nonumber \\ 
q_-&=&p_+- \frac{ \delta}{2w}  \\ 
q_+&=&\frac{1-w p_+}{W-w}+\frac{\delta}{2(W-w)}  \quad . \nonumber
\end{eqnarray}
Further we get the equation
\begin{equation}
g'(p_+)-g'(p_-)+g'(q_+)-g'(q_-)=0\quad.
\label{impo}
\end{equation}
Since  $g$ is concave $g'$ is monotonically  decreasing
and $g'(p_+)-g'(q_-)>0$ and $g'(q_+)-g'(p_-)>0$. 
Thus Eq. (\ref{impo}) has no solution, meaning  
that there is no extremum with $p_\pm$ and $q_\pm$ in $(0,1)$, and extrema are at the boundaries. 
The possibilities are $p_+=1$ or $p_-=0$, then $q_+=1$ and $q_-=0$.
Only $p_+=1$ or $p_-=0$ are compatible with the assumption that $S[p]>S[q]$; 
$p_+=1$ is only a special case of $p_-=0$ with $n=1$. Since $g(0)=0$ this immediately leads to the inequality
\begin{eqnarray}
\frac{|S_g[p]-S_g[q]|}{S_{\max}} & \leq &  (1-\phi)\frac{g\left(\frac{\delta}{2(1-\phi)W}\right)}{g(\frac 1W)}  \nonumber \\
&+& \phi\left|\frac{g\left(\frac{1}{\phi W}\right)}{g(\frac 1W)} -\frac{g\left(\frac{1-\frac{\delta}{2}}{\phi W}\right)}{g(\frac 1W)}\right|, 
\label{ineq1}
\end{eqnarray}
where $\phi=w/W$ is chosen such that the right hand side of the equation is maximal. 
Obviously, for any finite $W$ the right hand side can always be made as small as needed by choosing $\delta>0$
small enough. 
Now take the limit $W\to\infty$. If $f$ is continuous 
\begin{eqnarray}
&&\frac{|S_g[p]-S_g[q]|}{S_{\max}} \leq  \nonumber\\
&\leq& (1-\phi)\left(\frac{\delta}{2(1-\phi)}\right)^c  
+ \phi\left|\left(\frac{1}{\phi}\right)^c -\left(\frac{1- \frac{\delta}{2} }{\phi}\right)^c\right| \nonumber\\ 
&\leq& (1-\phi)^{1-c}\delta^c 
+ \phi^{1-c}\left|1 -\left(1- \frac{\delta}{2} \right)^c\right|\nonumber\\
&\leq& \delta^c + \left|1 -\left(1-c \frac{ \delta}{2} \right)\right| 
\leq \delta^c + \delta\quad. 
\label{ineq2}
\end{eqnarray}
Lesche-stability of $S_g$ follows  
since the right hand side of Eq. (\ref{ineq2}) can be made smaller than any given $\ep>0$ by choosing
$\delta>0$ small enough. This completes the first direction of the proof. 

If, on the other hand, $S_g$ is not Lesche-stable 
then there exists an $\ep>0$, such that $|S_g[p]-S_g[q]|/S_{\max}\geq \ep$, $\forall N$, 
implying  
\begin{equation}
(1-\phi)f\left(\frac{\delta}{2(1-\phi)}\right)   
+ \phi\left|f\left(\frac{1}{\phi}\right) -f\left(\frac{1-\frac{ \delta}{2} }{\phi}\right)\right|\geq \ep,
\label{ineq3}
\end{equation}
$\forall \delta>0$.
This again means that either $f(z)$ is discontinuous at $z=1/\phi$, or $\lim_{z\to0}f(z)>0$. 
Since $g(0)=0$ implies that $f(0)=0$, $f(z)$ has to be discontinuous at $z=0$. 
\end{proof}



{\bf Remark on higher order scaling exponents:} 

In principle the scheme of finding scaling exponents can be iterated to higher orders,
 i.e. to find a sequence of exponents
$(c,d,d_2,d_3,\dots)$. 
To see this define $\log^{(n+1)}(x)=\log(\log^{(n)}(x))$, with $\log^{(1)}(x)=\log(x)$ and
$\exp^{(n+1)}(x)=\exp(\exp^{(n)}(x))$, with $\exp^{(1)}(x)=\exp(x)$, then up to 
a precision $M$ 
\begin{equation} 
g(x)\sim x^c\prod_{m=1}^M \log^{(m)}\left(\frac{1}{x}\right) \quad, 
\end{equation} 
for small $x$.
The $n$'th scaling exponent $d_n$ can be determined  by
\begin{equation} 
\lim_{x\to 0}\frac{g\left(\lambda_n(x)x\right)}{\lambda_n(x)^{c}g(x)}
\prod_{m=1}^{n-1}
\left( \frac{\log^{(m)}\left(\frac{1}{x\lambda_n(x)}\right)
}{\log^{(m)}\left(\frac{1}{x}\right)} \right)^{-d_m}
= (1+a_n)^{d_n} , 
\end{equation} 
where the generalized scaling factor reads
\begin{equation} 
\lambda_n(x)^{-1}=
x\exp^{(n)}\left((1+a_n)\log^{(n)}\left(\frac{1}{x}\right)\right) \quad.
\end{equation} 
Note that this is a nested scheme of sequentially incorporating more information with higher levels,
i.e. if  $g\in(c,d,d_2,\dots, d_{n+1} )$, then also $g\in (c,d,d_2,\dots,g_n)$.
Using higher order asymptotic exponents makes it difficult to find a concave  representative
$g_{c,d,d_2,\dots,d_M}(x)$  and $g$ quickly becomes experimentally inaccessible.  
For macroscopic systems there may exist an upper limit $M$ such
that measuring $d_m$ for $m>M$ becomes senseless. 
For example, consider a system with $N\sim10^{23}$ particles and $\Omega\propto e^{N}$
states. Then $x\propto 1/\Omega$ and it does not make any sense
to measure $d_6$ or higher  because $\log^{(6)}(x)\sim -1.13$ is already negative, 
and $\log^{(7)}(x)$ is no longer defined on the principal branch of the logarithm.   

\end{document}